\documentclass[lettersize,journal]{IEEEtran}
\usepackage{amsmath,amsfonts,amssymb}
\usepackage{algorithmic}
\usepackage{algorithm}
\usepackage{array}
\usepackage[caption=false,font=normalsize,labelfont=sf,textfont=sf]{subfig}
\usepackage{textcomp}
\usepackage{stfloats}
\usepackage{url}
\usepackage{verbatim}
\usepackage{graphicx}
\usepackage{cite}
\usepackage{multirow}
\usepackage{multicol}
\usepackage{bm}

\begin{document}

\title{A Survey of Multi-sensor Fusion Perception for Embodied AI: Background, Methods, Challenges and Prospects}

\author{Shulan Ruan, Rongwei Wang, Xuchen Shen, Huijie Liu, Baihui Xiao, Jun Shi,\\ Kun Zhang, Zhenya Huang, Yu Liu, Enhong Chen, You He
\thanks{Shulan Ruan, Rongwei Wang, Xuchen Shen, Baihui Xiao, Yu Liu, and You He are with Tsinghua University~(e-mail: slruan@sz.tsinghua.edu.cn)}
\thanks{Huijie Liu, Jun Shi, Zhenya Huang, and Enhong Chen are with University of Science and Technology of China}
\thanks{Kun Zhang is with  Hefei University of Technology}
}



\maketitle

\begin{abstract}

Multi-sensor fusion perception~(MSFP) is a key technology for embodied AI, which can serve a variety of downstream tasks~(\textit{e.g.}, 3D object detection and semantic segmentation) and application scenarios~(\textit{e.g.}, autonomous driving and swarm robotics). 
Recently, impressive achievements on AI-based MSFP methods have been reviewed in relevant surveys.
However, we observe that the existing surveys have some limitations after a rigorous and detailed investigation. 
For one thing, most surveys are oriented to a single task or research field, such as 3D object detection or autonomous driving. Therefore, researchers in other related tasks often find it difficult to benefit directly.
For another, most surveys only introduce MSFP from a single perspective of multi-modal fusion, while lacking consideration of the diversity of MSFP methods, such as multi-view fusion and time-series fusion.
To this end, in this paper, we hope to organize MSFP research from a task-agnostic perspective, where methods are reported from various technical views.
Specifically, we first introduce the background of MSFP. Next, we review multi-modal and multi-agent fusion methods. A step further, time-series fusion methods are analyzed. In the era of LLM, we also investigate multi-modal LLM fusion methods. Finally, we discuss open challenges and future directions for MSFP.
We hope this survey can help researchers understand the important progress in MSFP and provide possible insights for future research.
\end{abstract}


\section{Introduction}
\label{s:introduction}

\IEEEPARstart{I}{n} recent years, benefiting from the rapid development of deep learning and large language model~(LLM), artificial intelligence~(AI) has achieved remarkable progress in various fields~\cite{ruan2021dae,ruan2025cpws,ruan2024color}.
As an important direction of AI, embodied AI refers to an intelligent form that uses physical entities as carriers and realizes autonomous decision-making and action capabilities through real-time perception in dynamic environments. 
Embodied AI has a wide range of application scenarios, such as autonomous driving and robot swarm intelligence~\cite{ren2024embodied,gupta2021embodied}. It has become an important research topic in the AI community in recent years,
which is also becoming a key path to break through the bottleneck of AI development and realize artificial general intelligence~(AGI).

In the construction of embodied AI systems, sensor data understanding is the core link between the physical world and digital intelligence. Different from the traditional vision-dominated perception mode, embodied agents need to integrate multi-modal sensor data to achieve a panoramic perception of the environment, such as visual cameras, millimeter-wave radars, LiDARs, infrared cameras and IMUs.
Multi-sensor fusion perception~(MSFP) is crucial to achieving robust perception and accurate decision-making capabilities of embodied AI. For example, visual cameras are easily disturbed by changes in illumination, and the performance of LiDAR will be greatly attenuated in rainy and foggy weather.

\begin{figure}[t]
    \centering
    \includegraphics[width=88mm]{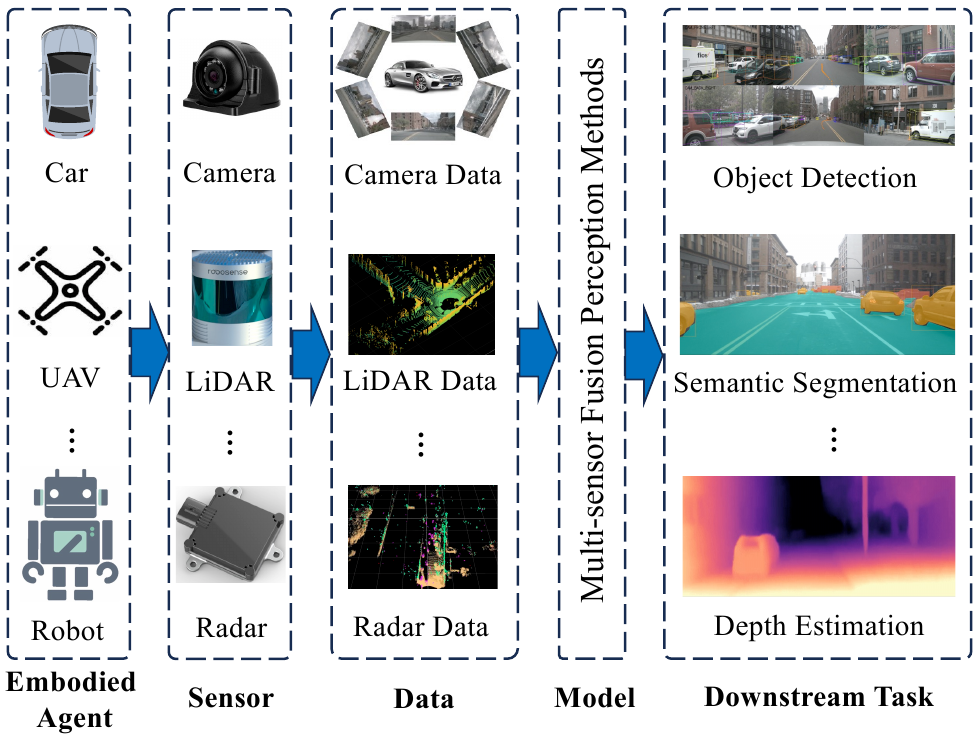}
    \caption{Overview of multi-sensor fusion perception pipeline.}
    \vspace{-4mm}
    \label{f:MSFP_pipeline}
\end{figure}

As shown in Fig.~\ref{f:MSFP_pipeline}, current research on multi-sensor fusion perception for embodied intelligence is mainly based on the paradigm of ``\texttt{Agent-Sensor-Data-Model-Task}". 
Existing multi-sensor fusion perception methods have achieved impressive success in many fields such as autonomous driving and industrial robots, but their application in embodied AI still faces some inherent challenges. Specifically, first, the heterogeneity of cross-modal data makes it difficult to unify the feature space. 
Second, the spatiotemporal asynchrony between different sensors may cause fusion errors. 
In addition, sensor failure~(\textit{e.g.}, lens contamination and signal obstruction) may cause dynamic loss of multi-modal information.

Around these issues, as shown in Table~\ref{t:surveys}, in recent years, some existing surveys have summarized the corresponding methods systematically~\cite{wangy2023multi,wangl2023multi,xiang2023multi,zhu2023camera,tang2023multi,du2024advancements,bin2024survey,song2024robustness,han2025multimodal}. 
Despite the impressive efforts, we observe that existing surveys still have some limitations after a detailed investigation. 
For one thing, most surveys are oriented to a single task or research field, such as 3D object detection or autonomous driving. Therefore, researchers in other related tasks often find it difficult to benefit directly.
For another, most surveys only introduce MSFP from a single perspective of multi-modal fusion, while lacking consideration of the diversity of MSFP methods, such as multi-agent fusion and time-series fusion.

\begin{table}[t]
    \caption{Overview of related surveys for MSFP.}
    \label{t:surveys}
    \centering
    \begin{tabular}{m{1.8cm}|m{0.5cm}m{2.4cm}m{2.2cm}}
    \hline
    \textbf{Survey} & \textbf{Year}  & \textbf{Field} &\textbf{Task}  \\ \hline
   Wang \textit{et al.}~\cite{wangy2023multi}& 2023 &Autonomous driving &3D object detection \\
   Wang \textit{et al.}~\cite{wangl2023multi}& 2023 &Autonomous
driving  &3D object detection\\ 
   Xiang \textit{et al.}~\cite{xiang2023multi}& 2023  &Autonomous driving &Non-specific\\ 
   Zhu \textit{et al.}~\cite{zhu2023camera}& 2023 &Non-specific  &SLAM\\ 
   Tang \textit{et al.}~\cite{tang2023multi}& 2023 &Autonomous driving  &3D object detection\\ 
   Du \textit{et al.}~\cite{du2024advancements}& 2024 &Non-specific &SLAM\\ 
   Bin \textit{et al.}~\cite{bin2024survey}& 2024 &Humanoid robot &Non-specific\\ 
   Song \textit{et al.}~\cite{song2024robustness}& 2024 &Autonomous driving &3D object detection\\ 
   Han \textit{et al.}~\cite{han2025multimodal}& 2025 &Robot &Non-specific\\ \hline
    \end{tabular}
\end{table}

To this end, in this paper, we hope to organize MSFP research from a task-agnostic perspective, where methods are reported purely from various technical views.
Specifically, we first introduce the background of MSFP, including various perception tasks, different sensor data, popular datasets, and corresponding evaluation criteria.
Next, we review the multi-modal fusion methods from point-level, voxel-level, region-level and multi-level fusion.
Along this line, we study the multi-agent fusion methods focusing on collaborative perception among multiple embodied agents and infrastructure.
A step further, time series fusion methods are also investigated, which fuse time-series~(\textit{e.g.}, several previous frames) sensor data to make predictions.
In the era of large models, we investigated MM-LLM fusion methods from vision-language based and vision-Lidar based methods, which are rarely included in previous surveys.
Finally, we discuss open challenges and future opportunities in MSFP comprehensively from data level, model level and application level.
We hope this survey can help researchers understand the important progress in MSFP in the past decade and provide possible insights for future research.

The remainder of this paper is organized as follows.
In Section~\ref{s:background}, we describe the background of MSFP from different sensor data, available datasets and various perception tasks.
In Section~\ref{s:multi-modal}, we introduce multi-modal fusion methods from different levels, \textit{e.g.}, point-level, voxel-level, region-level and multi-level.
In Section~\ref{s:multi-view}, we summarize multi-agent collaborative perception methods.
In Section~\ref{s:time series}, we review time-series fusion methods for MSFP.
In Section~\ref{s:MM-LLM}, we investigate current MM-LLM based methods for MSFP.
In Section~\ref{s:open challenges}, we discuss the open challenges and future directions in MSFP.
Finally, we conclude our work in Section~\ref{s:conclusion}.
\section{Background}
\label{s:background}

In this section, we focus on introducing the background of MSFP. First, we investigate some common sensors and their data types. 
Next, we collate relevant benchmark datasets available for researchers. 
Finally, we elaborate on various downstream tasks for MSFP.

\subsection{Sensor Data}

\subsubsection{Camera Data}
Cameras can capture rich appearance features of objects including colors, shapes and textures, which are crucial for various perception tasks. However, as passive sensors, cameras are sensitive to lighting conditions. Image quality deteriorates significantly at night and in adverse weather like fog and rain.

\subsubsection{LiDAR Data}
LiDAR calculates object distances by measuring the time difference between transmitted and received laser signals. 
It directly outputs high-precision 3D point clouds containing spatial geometric information, which has unique advantages in 3D perception. However, it is usually weather-sensitive. Due to the inherent sparsity and non-uniformity, effectively representing and understanding LiDAR point cloud data also remains challenging.

\subsubsection{MmWave Radar Data}
Millimeter-wave radars detect objects by transmitting and receiving radio waves. 
Compared to LiDAR point clouds, radar point clouds are sparser and struggle to accurately describe object contours. However, radars maintain good performance in adverse weather and can directly measure object velocities.

\subsection{Datasets}

\begin{table}[t]
\centering
\caption{Statistics of popular datasets for MSFP. Here, L, R, C denote LiDAR, Radar and Camera, respectively. U, S, H denote Urban, Suburban and Highway, respectively.}
\label{tab:3Ddatasets}
\begin{tabular}{l|lllll} 
\hline
\textbf{Dataset} & \textbf{Modality} & \textbf{Scenario} & \textbf{\# Class} & \textbf{\# Frame} \\ 
\hline
KITTI~\cite{geiger2012kitti}  & L + C & U + S + H & 8 & 15K \\ 
nuScenes~\cite{caesar2020nuscenes}  & L + R + C & U + S & 23 & 1M+ \\ 
Waymo Open~\cite{sun2020waymo}  & L + R + C & U + S & 4 & 1M+ \\ 
Cityscapes 3D~\cite{Gahlert2020Cityscapes3D} & C  & U & 30 & 35K \\ 
Argoverse~\cite{chang2019argoverse} & L + R + C & U & 15 & 1.75M+ \\ 
A*3D~\cite{pham2020a3d} & L + C  & U & 10 & 100K \\ 
ApolloScape~\cite{huang2018apolloscape} & L + C & U + S + H & 35 & 140K \\ 
AIODrive~\cite{weng2020all} & L + R + C  & U & 20 & 500K+ \\ 
H3D~\cite{patil2019h3d}  & L + C & U & 8 & 60K \\ 
\hline
\end{tabular}
\end{table}

\subsubsection{KITTI}
KITTI~\cite{geiger2012kitti} consists of $14,999$ images and corresponding point clouds, with $7,481$ for training and $7,518$ for testing. Annotations span eight categories and are classified into simple, medium, and difficult based on size, occlusion, and truncation levels.
The data collection vehicle was equipped with two grayscale cameras, two color cameras, a Velodyne 64-line LiDAR, four optical lenses, and a GPS system. Data was collected from approximately $50$ scenes across Karlsruhe and nearby German cities, covering urban, rural, and highway.

\subsubsection{nuScenes}
NuScenes~\cite{caesar2020nuscenes} was collected in Boston and Singapore. It includes $700$ training scenes, $150$ validation scenes, and $150$ test scenes. Each scene lasts approximately $20$ seconds with $40$ samples, totaling $5.5$ hours. The dataset contains $1.4$ million camera images, $390k$ LiDAR scans, $1.4$ million radar scans, and $1.4$ million annotated bounding boxes in $40k$ keyframes.
It is equipped with $6$ cameras covering 360-degree vision, a 32-beam LiDAR with $1.39$ million points per frame, $5$ millimeter-wave radars, and an inertial navigation system with GPS and IMU.

\subsubsection{Waymo Open}
Waymo Open~\cite{sun2020waymo} comprises perception and motion datasets.
Annotations in the perception dataset include $1.26$ million 3D bounding boxes, $1.18$ million 2D bounding boxes, panoptic segmentation labels for $100k$ images, $14$ keypoint annotations, and 3D semantic segmentation labels. 
The motion dataset contains $103,354$ clips with object trajectories.
The dataset includes daytime, nighttime, dawn, dusk, and rainy scenarios, but lacks extreme weather instances.

\subsubsection{Cityscapes 3D}
Cityscapes 3D~\cite{Gahlert2020Cityscapes3D} originated from the Cityscapes~\cite{Cordts2016Cityscapes} dataset, with the addition of 3D bounding box annotations. 
It consists of $5k$ finely annotated images (\textit{i.e.}, $2048\times 1024$ pixels) and 20k coarsely annotated images, which are employed for 3D scene understanding tasks in urban streetscapes, \textit{e.g.}, instance-level semantic segmentation.

\subsubsection{Argoverse}
Argoverse~\cite{chang2019argoverse} was collected with two 32-channel LiDAR sensors, seven surround-view cameras, and two forward stereo cameras, covering 360 degrees. 
It encompasses a 3D tracking dataset with 1k 3D annotated samples, covering 30 object categories. 
It also encompasses a motion prediction dataset with $250k$ samples providing trajectory data for scenes, $20k$ unlabeled LiDAR data sample, and $1k$ high-definition maps, which offer abundant semantic annotation information regarding road infrastructure and traffic rules.

\subsubsection{A*3D}
A*3D~\cite{pham2020a3d} was mainly collected on the urban roads of Singapore. It contains over 39k annotated frames, each of which is labeled with 2D and 3D bounding boxes and includes a target tracking ID across frames. The A*3D dataset encompasses different sensor data such as high-density 3D point clouds, high-definition RGB images, and IMU data, covering a 360-degree view. It covers various weather conditions such as daytime, nighttime, and rainy days, as well as acquisition scenarios of different urban road conditions.

\subsubsection{ApolloScape}
ApolloScape~\cite{huang2018apolloscape} was collected by two LiDAR sensors, six video cameras, and an IMU/GNSS system, encompassing over $140k$ high-resolution images. It covers multiple time periods and weather conditions and has a total of $25$ categories. 

\subsubsection{AIODrive}
AIODrive~\cite{weng2020all} was developed by a research team of Carnegie Mellon University, and it is also targeted at the urban scene. The synthetic sensor data of AIODrive encompasses five $1920 \times 720$ RGB cameras and five depth cameras, one radar, one Velodyne-64 LiDAR, an IMU, GPS, and three long-range high-density LiDARs.

\subsubsection{H3D}
H3D~\cite{patil2019h3d} mainly focuses on 3D object detection and tracking in urban environments. It offers approximately 160 urban scenes by means of Velodyne HDL-64E LiDAR and three high-resolution RGB cameras, totaling approximately $27k$ frames. Each frame contains detailed 3D bounding boxes and tracking identity information of objects.

\subsection{Perception Tasks}
\subsubsection{Object Detection}
Object detection is one of the most fundamental tasks in extensive perception systems, whose core goal is to accurately locate and identify various types of objects through the data obtained by sensors. 
In 2D object detection, the system needs to output the category information of the object and the 2D bounding box represented by $(x, y, w, h)$. 
In the 3D object detection scenario, the detection results need to include the 3D position coordinates $(x, y, z)$, 3D size information $(l, w, h)$ and heading angle $\theta$ of the target.

\subsubsection{Semantic Segmentation}
The semantic segmentation task aims to classify each basic unit in a scene, such as an image pixel, into semantic categories. 
Specifically, given a set of input data (\textit{e.g.}, a set of image pixels $\bm{I} = \{\bm{I}_1, \bm{I}_2, ..., \bm{I}_n\}$), and a predefined set of semantic categories $\bm{y} = \{y_1, y_2, ..., y_k\}$, the segmentation model needs to assign a corresponding semantic label or class probability distribution to each basic unit $\bm{I}_i$. 

\subsubsection{Depth Estimation}
Depth estimation aims to obtain the depth information of the scene from the sensor data and provide 3D geometric understanding for embodied agents. Given an input image $\bm{I} \in \mathcal{R}^{M\times N}$ and the corresponding sparse depth map $\bm{D}_{s} \in \mathcal{R}^{M\times N}$, the depth estimation system needs to output a dense depth map $\bm{D}_{d}$, where the depth complement process can be represented as the mapping function $\bm{D}_d=f(\bm{I}, \bm{D}_{s})$. Through depth estimation, the system is able to obtain accurate 3D position information of objects in the scene, which is critical for downstream tasks such as path planning and decision control.

\subsubsection{Occupancy Prediction}
Occupancy prediction can provide an intensive semantic understanding of 3D space. By discretizing a continuous 3D space into voxels, the occupancy perception model can predict the occupancy state and semantic category of each voxel, thus providing a complete scene representation for autonomous decisions.

\section{Multi-modal Fusion Methods}
\label{s:multi-modal}

\begin{table}[t]
    \centering
    \caption{Categorization of multi-modal fusion methods for MSFP.}
    \label{t:multi-modal}
    \scalebox{0.9}{\begin{tabular}{m{1.4cm}|m{3.5cm}|m{3.5cm}}
       \hline
       \textbf{Category}  & \textbf{Feature} & \textbf{Methods} \\ \hline
         Point-level fusion& Integrate geometric coordinate info of LiDAR points with semantic details of images at individual point level. & PointFusion~\cite{xu2018pointfusion}, PointPainting~\cite{vora2020pointpainting}, MVP~\cite{yin2021multimodal}, DeepFusion~\cite{li2022deepfusion} \\ \hline
         Voxel-level fusion& Convert irregular LiDAR point clouds to regular voxel grids for efficient processing, preserving geometric information while integrating multi-modal data to enhance semantic richness. &CenterFusion~\cite{nabati2021centerfusion}, PointAugmenting~\cite{wang2021pointaugmenting},  UVTR~\cite{li2022unifying}, SFD~\cite{wu2022sparse} \\ \hline
         Region-level fusion& Aggregate region-specific features across modalities via spatial alignment, leveraging inter-modal spatial correspondence with late-stage fusion. & AVOD~\cite{ku2018joint}, RoarNet~\cite{shin2019roarnet}, AR-CNN~\cite{zhang2019weakly}, E2E-MFD~\cite{zhang2024e2e}\\ \hline
         Multi-level fusion& Combine multi-modal fusion across multiple hierarchical levels, using techniques like multi-stage fusion, attention, or contrastive learning to enhance perception robustness. & MVX-Net~\cite{sindagi2019mvx}, RCBEV~\cite{zhou2023bridging}, MBNet~\cite{zhou2020improving}, CSSA~\cite{cao2023multimodal}\\ \hline
    \end{tabular}}
\end{table}

By fusing data from multi-modal sensors, embodied agents can reduce perception blind spots and achieve more comprehensive environmental perception.
For example, LiDAR can provide accurate depth information, while cameras retain more detailed semantic information.
Therefore, how to better fuse the multi-modal data from different sensors to provide more accurate and robust perception has become a hot research topic in extensive applications.
As shown in Table \ref{t:multi-modal}, in this section, we introduce various methods from different fusion levels, \textit{i.e.}, point-level, voxel-level, region-level and multi-level.

\begin{figure}[t]
    \centering
    \includegraphics[width=88mm]{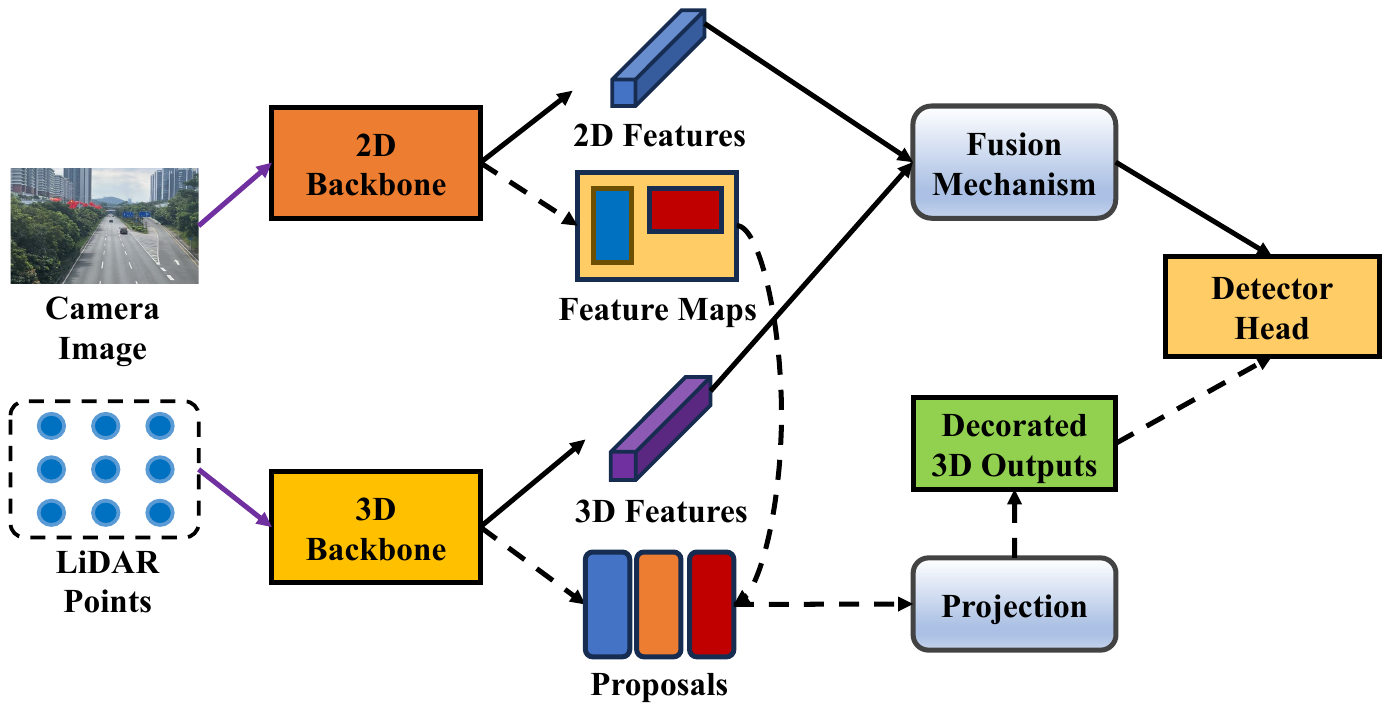}
    \caption{Overview of point-level fusion pipeline.}
    \label{f:point}
\end{figure}

\subsection{Point-level Fusion}
\label{ss:3.2}

Fig.~\ref{f:point} presents the typical pipeline of point-level fusion methods, which aim to achieve feature fusion at the individual point level between LiDAR point clouds and image data. By integrating the geometric coordinate information of point clouds with the semantic details of images (\textit{e.g.}, color and category attributes), the multi-modal perception accuracy can be enhanced.
Among various methods, PointNet~\cite{qi2017pointnet} and PointNet++~\cite{qi2017pointnet++} initially directly process point clouds without relying on other forms such as voxels. They were only used for LiDAR-based 3D object recognition. 
Frustum PointNets~\cite{qi2018frustum} extends PointNet by converting 2D candidate boxes into 3D frustums and performing segmentation and regression directly on raw point clouds. 
PointFusion~\cite{xu2018pointfusion} takes a more standard approach by extracting features separately from RGB images and point clouds using CNNs and PointNet, respectively, and then concatenating them for fusion. 

However, the initial fusion struggles to capture complex cross-modal relationships. PI-RCNN~\cite{xie2020pi} improves on this with a two-stage process, using attentive aggregation to refine the fusion of 3D proposals and 2D semantic features, allowing for more detailed processing.
Methods like PointPainting~\cite{vora2020pointpainting} and FusionPainting~\cite{xu2021fusionpainting} annotate each LiDAR point with image features, with the former projecting LiDAR points onto segmentation masks and the latter using adaptive attention for semantic-level fusion. These approaches better handle point cloud sparsity compared to proposal-first methods like PI-RCNN. Similarly, MVP~\cite{yin2021multimodal} enhances sparse point clouds by projecting 2D detection results into virtual 3D points and merging them with LiDAR data, compensating for the limitations of LiDAR in detecting small or distant objects. 
DeepFusion~\cite{li2022deepfusion} employs a cross-attention mechanism to dynamically align LiDAR and image features and resolves geometric misalignment problems through reverse data augmentation. GraphAlign~\cite{song2023graphalign} further optimizes the alignment process using graph-based feature matching. It achieves pixel-level precise matching between point cloud geometric features and image semantic features through the graph feature alignment and self-attention feature alignment modules, solving the core problems of inaccurate geometric positions and ambiguous semantic associations in multi-modal fusion.

\subsection{Voxel-level Fusion}
\label{ss:3.3}

\begin{figure}[t]
    \centering
    \includegraphics[width=80mm]{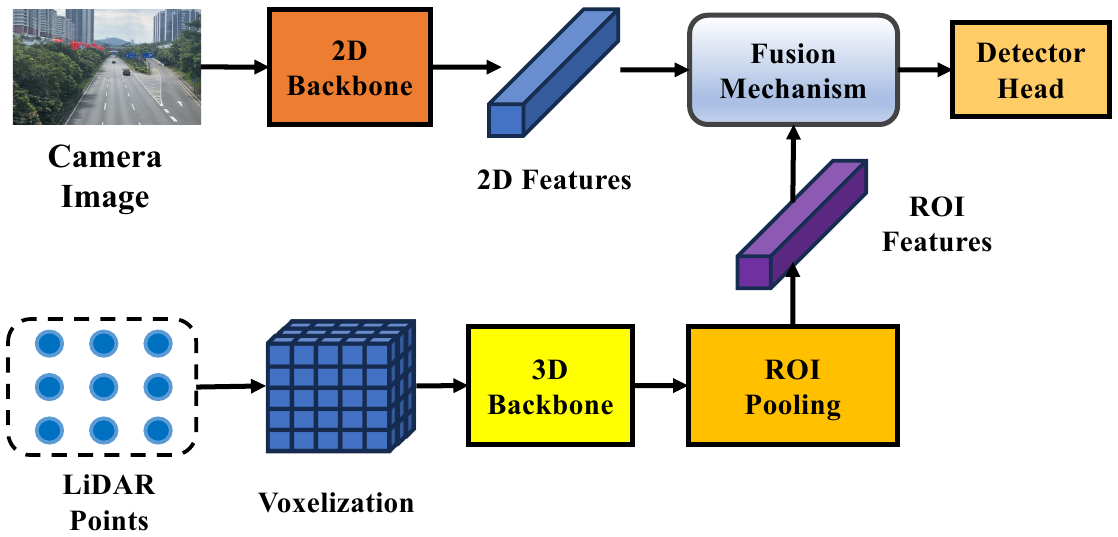}
    \caption{Overview of voxel-level fusion pipeline.}
    \label{f:voxel}
\end{figure}

Voxel-level fusion methods convert irregular LiDAR point clouds into regular grids (\textit{e.g.}, voxels or pillars), enabling efficient processing while preserving geometric information. Fig.~\ref{f:voxel} shows a typical pipeline of voxel-level fusion methods. 
To leverage the semantic richness of images, camera images are integrated into voxel-based methods for better perception ability, especially in sparse or occluded scenarios.

To address issues like inaccurate height information, CenterFusion~\cite{nabati2021centerfusion} expands radar points into 3D pillars, associates radar detections with image objects. However, voxel-level methods often suffer from ``feature blurring" due to spatial information loss within voxels. VPFNet~\cite{zhu2022vpfnet} mitigates this by using a voxel-RoI pooling layer and virtual points to align and aggregate features from LiDAR and images. PointAugmenting~\cite{wang2021pointaugmenting} enhances LiDAR points with image features and voxelizes the augmented point cloud. However, projecting 3D points to the image plane can degrade performance in occluded regions. VFF~\cite{li2022voxel} introduces a point-to-ray projection method, fusing image features along rays to provide richer contextual information, which is particularly beneficial for detecting occluded and distant objects.

For feature alignment, AutoAlign~\cite{chen2022autoalign} introduces a learnable multi-modal fusion framework, dynamically aligning image and point cloud features without relying on a projection matrix. As the enhanced version of AutoAlign, AutoAlignV2~\cite{chen2022deformable} uses deformable attention and sparse sampling to improve efficiency and reduce computational costs, while simplifying data augmentation compared to earlier methods like PointAugmenting. VoxelNextFusion~\cite{song2024voxelnextfusion} combines voxel features with corresponding and surrounding pixel features using self-attention for point and block fusion. This approach effectively resolves resolution mismatches and improves the detection of long-range and challenging objects. 

\subsection{Region-level Fusion}
\label{ss:3.4}

Region-level fusion methods focus on aggregating region-specific information, such as feature maps, ROIs, or region proposals, from 2D images and other modalities. These methods are particularly effective in scenarios where spatial alignment between modalities is easier to achieve.
AVOD~\cite{ku2018joint} introduced a multi-modal fusion region proposal network that processes BEV and RGB images separately to generate high-resolution feature maps. By regressing direction vectors, AVOD resolves ambiguity in direction estimation. Similarly, RoarNet~\cite{shin2019roarnet} employs a two-stage framework, where the first stage predicts 3D poses directly from images to avoid projection-related information loss, and the second stage refines these predictions using point cloud reasoning. TransFusion~\cite{bai2022transfusion} leverages Transformers for LiDAR-camera fusion by establishing soft associations between LiDAR points and image pixels. It can adapts to contextual information, addressing robustness issues caused by poor image quality or sensor calibration errors. 

\begin{figure}[t]
    \centering
    \includegraphics[width=80mm]{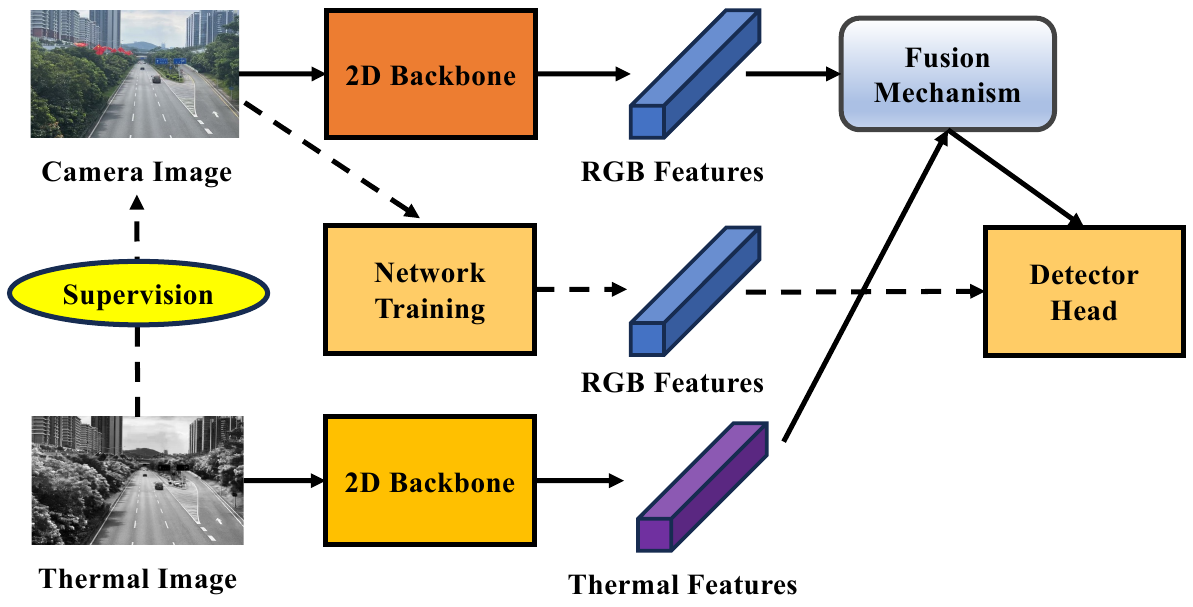}
    \caption{Overview of region-level fusion pipeline.}
    \label{f:region}
\end{figure}

For thermal-RGB fusion, region-level methods are more common due to simpler feature alignment.
Fig.~\ref{f:region} reflects the pipeline of region-level fusion methods. CMT-CNN~\cite{xu2017learning} tackles pedestrian detection in low-light conditions by reconstructing thermal regions corresponding to RGB candidate regions, fusing cross-modal information via a multi-scale detection network. AR-CNN~\cite{zhang2019weakly} addresses misalignment between RGB and thermal images by predicting positional shifts and adaptively aligning regional features. Dual-stream architectures like GAFF~\cite{zhang2021guided} and RSDet~\cite{zhao2024removal} process RGB and thermal images separately before fusing their features. GAFF employs intra- and inter-modal attention mechanisms for feature selection, while RSDet refines fusion through redundant spectrum removal and dynamic feature selection.

\subsection{Multi-level Fusion}
\label{ss:3.5}

\begin{figure}[t]
    \centering
    \includegraphics[width=85mm]{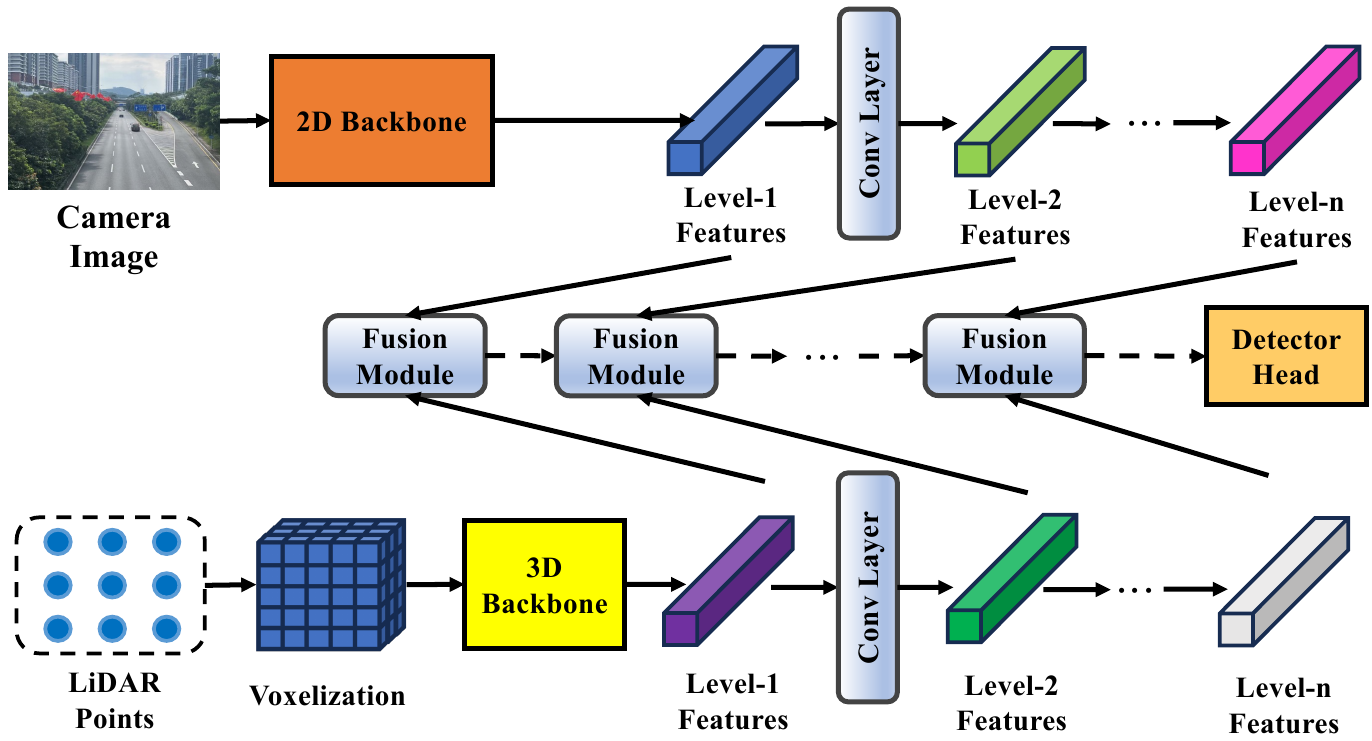}
    \caption{A multi-level fusion pipeline.}
    \label{f:multi}
\end{figure}

Multi-level fusion integrates multi-modal information from different levels to enable a more comprehensive perception. 
Fig.~\ref{f:multi} illustrates the pipeline for multi-level fusion methods.
In literature, Liang \textit{et al.}~\cite{liang2018deep} exploit continuous convolutions to fuse image and LIDAR feature maps at different levels in the BEV space.
Zhu \textit{et al.}~\cite{zhu2021cross} present a two-stage cross-modality fusion method to enhance semantic richness and local proposal representation from both point level and region level. In this way, perception performance can be improved under sparse and occluded scenarios.
Similarly, MVX-Net~\cite{sindagi2019mvx} performs point-level and voxel-level fusion. MMF~\cite{liang2019multi} further extends the idea to a multi-task framework, such as 2D/3D detection, ground estimation, and depth completion.

To improve robustness, EPNet~\cite{huang2020epnet} introduces the LI-Fusion module to reduce irrelevant information interference by fusing image and point cloud features at different scales.
As an enhanced version, EPNet++~\cite{liu2022epnet++} further introduces bidirectional information interaction, which adopts point cloud features to refine image features and vice versa. 
In this way, EPNet++ can achieve a more robust feature representation. RCBEV~\cite{zhou2023bridging} focuses on dynamic object perception by bridging radar-camera feature differences, while 
DVF~\cite{mahmoud2023dense} enhances representations in low-density areas by generating multi-scale dense voxel features, avoiding noisy 2D predictions with 3D bounding box labels.
LoGoNet~\cite{li2023logonet} combines global and local fusion with dynamic feature aggregation to improve detection accuracy in complex environments.

Some recent methods employ attention mechanisms and contrastive learning to enhance multi-modal fusion. For example, CAT-Det~\cite{zhang2022cat} encodes global contextual information across modalities using contrastive learning. SeaDATE~\cite{dong2024seadate} uses dual attention and contrastive learning to extract deep semantic information, while CSSA~\cite{cao2023multimodal} employs lightweight channel switching and spatial attention for efficient fusion. Fusion-Mamba~\cite{dong2024fusion} resolves alignment issues with an improved Mamba structure and hidden state space. 

\section{Multi-agent Fusion Methods}
\label{s:multi-view}

\begin{figure}[t]
    \centering
    \includegraphics[width=85mm]{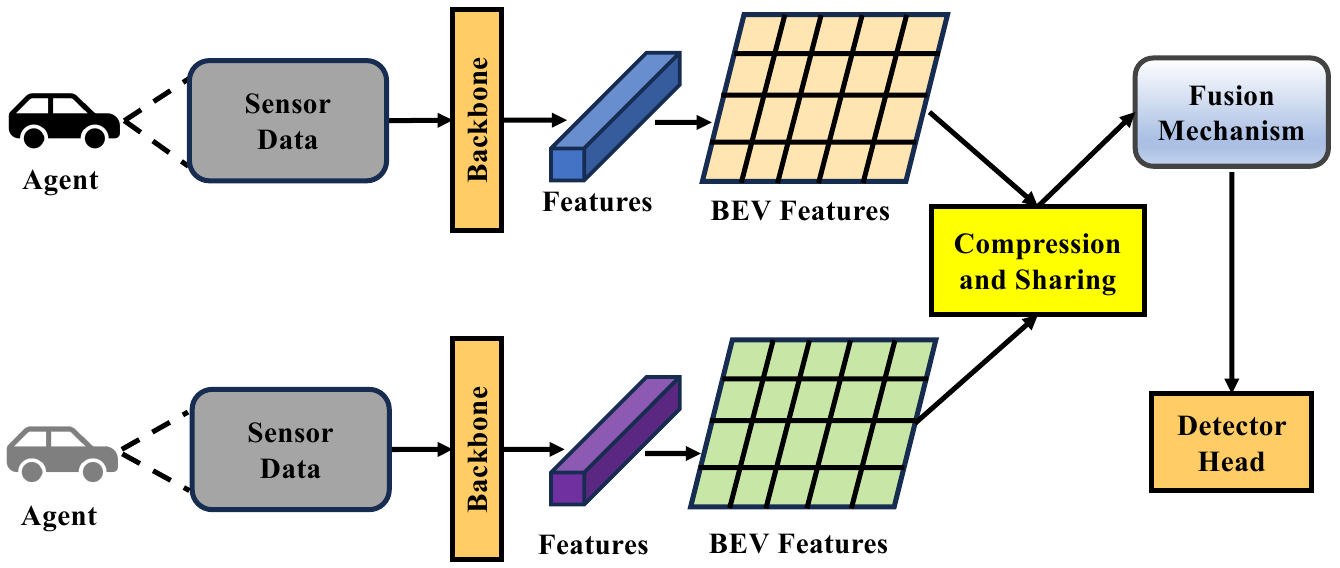}
    \caption{A simple agent-to-agent~(A2A) fusion pipeline.}
    \label{f:agent}
\end{figure}

In complex open environments, especially when visibility is obstructed or in adverse weather conditions, the perception system of a single embodied agent faces numerous challenges.
Collaborative perception technology can integrate perception data from multiple agents and infrastructure, which is crucial for addressing occlusion and sensor failure issues. 
In this section, we will focus on the multi-view fusion of agent-to-agent~(A2A) collaborative perception. 
Fig.~\ref{f:agent} shows a simple pipeline of A2A fusion.

CoBEVT~\cite{xu2022cobevt} is the first universal multi-agent multi-camera perception framework. It generates BEV segmentation predictions through sparse transformers for collaborative processing. CoBEVT incorporates an axial attention module to efficiently fuse multi-agent multi-view camera features, capturing both local and global spatial interactions. 
CoCa3D~\cite{hu2023collaboration} proposes an innovative collaborative camera-only framework. It addresses the issue of depth prediction bias by allowing multiple agents equipped only with cameras to share visual information. By sharing depth information at the same points, CoCa3D reduces errors, improves handling of depth ambiguities, and extends detection capabilities to occluded and long-range areas, which are typically challenging for single-agent systems. 
V2VNet~\cite{wang2020v2vnet} introduces a graph neural network-based framework for fusing intermediate feature representations from multiple vehicles. 
MACP~\cite{ma2024macp} explores efficient model adaptation using pre-trained single-agent models to achieve collaborative perception with low parameter count and communication costs. 
HM-ViT~\cite{xiang2023hm} proposes a unified framework for the multi-modal A2A perception problem, capable of fusing multi-view image and LiDAR point cloud features from different types of sensors, enabling efficient multi-modal cooperative perception. 
MRCNet~\cite{hong2024multi} addresses motion blur issues by introducing a motion enhancement mechanism that reduces the impact of motion blur on object detection by capturing motion context, achieving better performance in noisy scenarios. 

In addition, some works focus on improving communication issues in collaborative perception to achieve more efficient and robust cooperation. When2Com~\cite{liu2020when2com} proposed a framework for learning how to construct communication groups and when to communicate. By utilizing handshake mechanisms and asymmetric message sizes, it reduces bandwidth usage and achieves good performance in semantic segmentation and 3D shape recognition tasks. Who2Com~\cite{liu2020who2com} improves accuracy in semantic segmentation tasks by learning handshake communication mechanisms and uses less bandwidth compared to centralized methods. How2Com~\cite{yang2024how2comm} further proposed an information-theoretic communication mechanism and spatiotemporal collaborative Transformer, which improves collaborative perception by feature filtering, delay compensation, and spatiotemporal fusion, resulting in more efficient and robust cooperation in 3D object detection tasks. 
CodeFilling~\cite{hu2024communication} effectively optimizes the representation and selection of collaborative messages through an information-filling strategy and codebook compression techniques, achieving efficient collaborative perception with low communication costs.

\section{Time Series Fusion}
\label{s:time series}

\begin{figure}[t]
    \centering    
    \includegraphics[width=85mm]{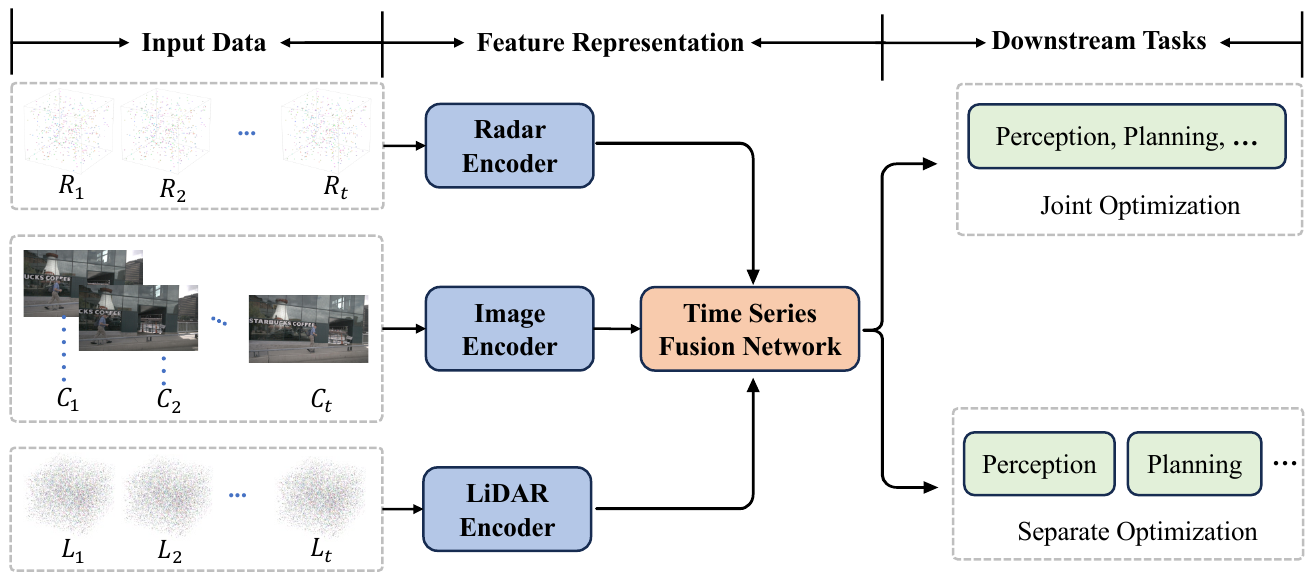}
    \caption{Framework overview of time series multi-sensor fusion network.}
    \label{f:time_series_frame}
\end{figure}

\begin{figure*}[t]
    \centering    
    \includegraphics[width=180mm]{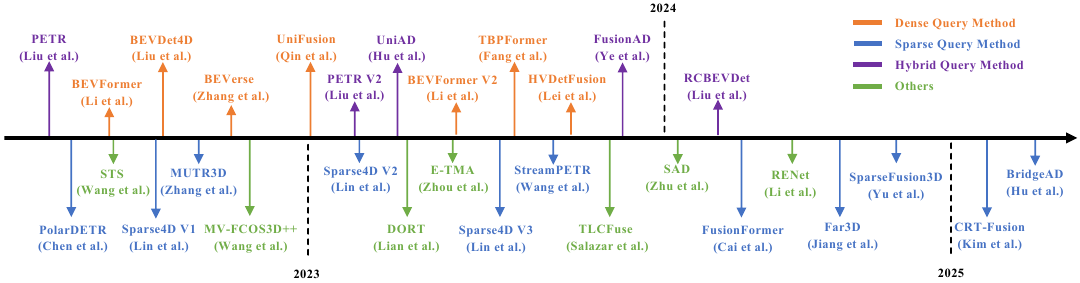}
    \caption{Timeline of time series fusion methods.}
    \label{f:time-series-timeline}
\end{figure*}

Time-series fusion represents a critical component in MSFP systems, addressing single-frame limitations and enhancing perceptual continuity across spatiotemporal domains. 
Fig.~\ref{f:time_series_frame} shows a simple pipeline of A2A fusion.
With the emergence of transformer architectures in computer vision, query-based fusion methods have become predominant, wherein perception features encoded as queries interact with spatiotemporal keys and values to achieve effective feature alignment. 
As shown in Table~\ref{t:time-series},
these methods can be taxonomized into three principal categories, \textit{i.e.}, dense query, sparse query and hybrid query.
Fig.~\ref{f:time-series-timeline} depicts the timeline of query-based methods for time series multi-sensor fusion.

\begin{table}[t]
    \centering
    \caption{Categorization of query-based time series fusion methods.}
    \label{t:time-series}
    \scalebox{0.96}{
    \begin{tabular}{m{1cm}|m{3.3cm}|m{3.3cm}}
       \hline
       \textbf{Category}  & \textbf{Feature} & \textbf{Methods} \\ \hline
         Dense query & Maintain fixed spatial positions in high-resolution representation spaces & BEVFormer~\cite{li2022bevformer}, BEVFormer v2~\cite{yang2023bevformer} \\ \hline
         Sparse query & Efficiently focus computational resources on regions of interest & StreamPETR~\cite{streampetr}, Sparse4D (v1~\cite{2211.10581}/v2~\cite{lin2023sparse4d}/v3~\cite{2311.11722}), SparseFusion3D~\cite{10314799}\\  \hline
         Hybrid query & Combine dense and sparse paradigms & UniAD~\cite{hu2023_uniad}, FusionAD~\cite{ye2023fusionad}, RCBEVdet~\cite{lin2024rcbevdet}\\ \hline
    \end{tabular}
    }
\end{table}

\subsection{Dense Query Methods}
Dense query methods assign a fixed rasterized spatial position to each query point within high-resolution 3D space or BEV space~\cite{10670223}. 
Among these methods, BEV-based Dense Frameworks are particularly representative, with BEVFormer~\cite{li2022bevformer} emerging as a seminal BEV perception model. Based on DETR~\cite{carion2020end} and Deformable DETR~\cite{zhu2021deformable}, BEVFormer achieves adaptive feature interaction in multiple camera views through deformable attention mechanisms. In contrast to the decoder in DETR3D~\cite{detr3d} which relies on sparse object queries, BEVFormer incorporates an additional encoder based on dense BEV queries to generate dense BEV features, facilitating semantic segmentation tasks. BEVFormer fuses temporal information between timestamps $t-1$ and $t$ through a Temporal Self-Attention module in its encoder, which can be interpreted as an implementation of the deformable attention mechanism~\cite{zhu2021deformable}. Building upon this foundation, BEVFormer v2~\cite{yang2023bevformer} adopts a two-stage detection architecture that integrates perspective view detection with BEV detection. This enables BEVFormer v2 to learn 3D scene representations adaptively through perspective supervision rather than relying on expensive depth pre-training data.

Based on LSS~\cite{philion2020lift}, a representative of depth-based bottom-up approaches, BEVDet4D~\cite{huang2022bevdet4d} extends 3D detection to the 4D temporal domain. BEVDet~\cite{huang2021bevdet} adheres to the LSS paradigm and proposes a framework for 3D detection of multi-view cameras in BEV. Moreover, BEVDet4D retains the BEV features generated from the previous frame and fuses them with the current frame features via spatial alignment and feature concatenation. To address the impact of vehicle self-motion, the authors proposed an ego-motion compensation method based on convolutions and ensured the accuracy of feature alignment through an auxiliary task. As a unified perception and prediction framework, BEVerse~\cite{zhang2022beverse} generates 4D BEV representations from multi-camera video sequences through shared feature extraction and lifting modules.

Furthermore, there exist some methods that possess unique architectural designs or have undergone optimization for specific tasks. Based on~\cite{li2022bevformer,yang2023bevformer,huang2021bevdet,huang2022bevdet4d}, UniFusion~\cite{qin2023unifusion} puts forward a unified temporal-spatial fusion framework, introducing the concept of virtual views. Historical frames are viewed as additional camera views with spatial transformation relationships, facilitating the parallel processing of temporal and spatial information.
On this basis, TBPFormer~\cite{fang2023tbp} focuses on a more general architectural design, proposing the PoseSync BEV Encoder for feature alignment and synchronization and designing a temporal pyramid transformer for multiscale feature extraction and future state prediction. HVDetFusion~\cite{lei2023hvdetfusion} additionally supports the radar modality and designs a two-stage decoupled detection architecture based on BEVDet4D~\cite{huang2022bevdet4d}. It utilizes a consecutive sequence of 16 frames (\textit{i.e.}, 8 frames of history, and 8 frames of future) for feature extraction and fusion, significantly enhancing the detection and speed estimation accuracy of moving objects.

\subsection{Sparse Query Methods}

In complex open environments, when temporal information is integrated into the fusion process, the volume of data that the network processes becomes more enormous. Furthermore, for those applications necessitating real-time decision-making tasks, model inference speed should be placed with stringent requirements. Consequently, sparse query methods, known for their efficiency, accuracy, and suitability for sparse perception tasks, have gained increasing popularity in the industry. 
In dense BEV feature construction methods, simply warping historical multi-frame BEV features to the current frame and concatenating them often achieves good results. However, in sparse query feature representations, this explicit fusion method becomes extremely challenging. Consequently, many approaches resort to having each query feature interact with multi-frame image features, which introduces substantial computational overhead. StreamPETR~\cite{streampetr} addresses this by systematically propagating long-term information across frames through object queries. This object-centric temporal modeling paradigm avoids the computational burden of modeling temporal relationships in dense BEV features.

Following StreamPETR, subsequent work has achieved further improvements in feature representation and sampling strategies. 
However, BEV methods face inherent trade-offs between perception range, accuracy, and computational efficiency, while being unable to directly perform 2D perception tasks in the image domain.
In response to these limitations, Sparse4D v1~\cite{2211.10581} achieves efficient spatio-temporal feature extraction through 4D keypoint sampling and hierarchical feature fusion. 
Based on Sparse4D v1, Sparse4D v2~\cite{lin2023sparse4d} adopts a recurrent approach using sparse instances for temporal information propagation, avoiding multi-frame sampling to improve feature fusion efficiency. 
Building upon this, Sparse4D v3~\cite{2311.11722} takes a step further by proposing temporal instance denoising and quality estimation, which simultaneously accelerates model convergence and enhances performance. 

Multi-task learning plays a crucial role in various perception tasks. While jointly training multiple tasks can make neural networks cumbersome, using sparse queries offers an elegant solution. 
MUTR3D~\cite{zhang2022mutr3d} is the first end-to-end 3D multi-target tracking framework that connects target detection with downstream tasks such as path planning and trajectory prediction through 3D MOT, and proposes a 3D track query mechanism that can model spatio-temporal consistency of targets across frames.
Based on MUTR3D, PF-Track~\cite{pang2023standing} adopts the ``tracking by attention" framework and represents tracked instances coherently over time with object queries. In the case of long-term occlusions, PF-Track maintains object positions and enables re-association through the Future Reasoning module, which digests historical information and predicts robust future trajectories up to 4 seconds ahead. 

In addition, recent research has shown an emerging trend towards exploring new paradigms for sparse multi-modal temporal fusion. Earlier methods like FusionFormer~\cite{cai2024fusionformer} focused on temporal fusion of BEV features for 3D object detection, utilizing deformable attention mechanisms and residual structures for feature alignment and fusion. 
Despite the intuition of dense BEV feature-based methods, most of them suffer from serious information loss when processing Z-axis information.
Also on the basis of DETR~\cite{wang2022detr3d}, QTNet~\cite{hou2023query} proposed a novel temporal fusion paradigm leveraging sparse queries. The motion-guided timing modeling (MTM) module effectively handles cross-modal correlation between point clouds and image features, achieving better performance while maintaining a lightweight architecture. SparseFusion3D~\cite{10314799} further advances this approach, introducing the MSPCP module to predict the point cloud offset and incorporating the radar-assisted query initialization strategy to deal with the sparsity challenge. The evolution from the MTM of QTNet to the MSPCP of the SparseFusion3D module represents a technological shift from simple feature alignment to explicit motion-based modeling. The SQS module of SparseFusion3D represents the development of multi-modal fusion strategies from feature concatenation to more sophisticated fusion strategies such as adaptive weighted fusion.
Furthermore, CRT-Fusion~\cite{kim2024crt} addresses the challenge of incorporating object motion in camera-radar temporal fusion by introducing multi-step motion queries that differentiate each future time step. The method employs a Motion Feature Estimator to predict pixel-wise velocity and a Motion Guided Temporal Fusion module that aligns features across timestamps in a recurrent manner, achieving superior performance by explicitly considering object dynamics.

\subsection{Hybrid Query Methods}
Hybrid query methods combine dense and sparse query paradigms to balance computational efficiency with comprehensive scene understanding. These approaches strategically employ sparse queries for object-level tasks while maintaining dense representations for spatial-complete tasks, achieving optimal performance across multiple perception objectives.

UniAD~\cite{hu2023planning} exemplifies this hybrid architecture by integrating perception, prediction, and planning in a unified framework. It utilizes sparse object queries for efficient detection and tracking, while maintaining dense BEV features for trajectory prediction and planning tasks. This dual representation enables comprehensive scene understanding without sacrificing real-time performance.
Building on the success of UniAD, FusionAD~\cite{ye2023fusionad} extends the hybrid approach to multi-modal temporal fusion. It processes camera and LiDAR data through a transformer-based architecture that adaptively switches between sparse and dense representations based on task requirements, demonstrating the flexibility of hybrid query methods in handling heterogeneous sensor data.

Multi-modal hybrid query methods, which effectively handle heterogeneous data from multiple sensors (different surrounding view video cameras or multi-modal sensors, \textit{e.g.}, 4D millimeter wave radar, LiDAR, camera) through elaborate architectures, demonstrate superior capabilities in temporal-spatial feature extraction and fusion. 
Taking inspiration from CRN~\cite{kim2023crn}, RCBEVdet~\cite{lin2024rcbevdet} introduces a dual-stream network. For the radar stream, RadarBEVNet which generating dense BEV features is designed for point cloud BEV feature extraction.
For the camera stream, an image backbone and view transformer from LSS~\cite{philion2020lift} are utilized for feature representation. 
Then, with a cross-attention multi-layer fusion module based on deformable DETR, effective 4D millimeter-wave radar-camera fusion can be better performed.

\section{MM-LLM Fusion Methods}
\label{s:MM-LLM}

\begin{table}[t]
    \centering
    \caption{Categorization of MM-LLM fusion methods for MSFP.}
    \label{t:MM-LLM}
    \begin{tabular}{m{2cm}|m{2.8cm}|m{2.8cm}}
       \hline
       \textbf{Category}  & \textbf{Feature} & \textbf{Methods} \\ \hline
        Visual-Language & Combine visual and textual data for semantic alignment. 
        & Scedrivex~\cite{zhao2025sce2drivex}, X-Driver~\cite{liu2025x}, Mpdrive~\cite{zhang2025mpdrive}, SafeAuto~\cite{zhang2025safeauto}\\ \hline
        Visual-LiDAR-Language & Integrate visual, LiDAR and language data for 3D spatial understanding. & DriveMLM~\cite{wang2023drivemlm}, MAPLM~\cite{cao2024maplm}, LiDAR-LLM~\cite{yang2025lidar} \\ \hline
    \end{tabular}
\end{table}

\begin{figure}[t]
    \centering
    \includegraphics[width=85mm]{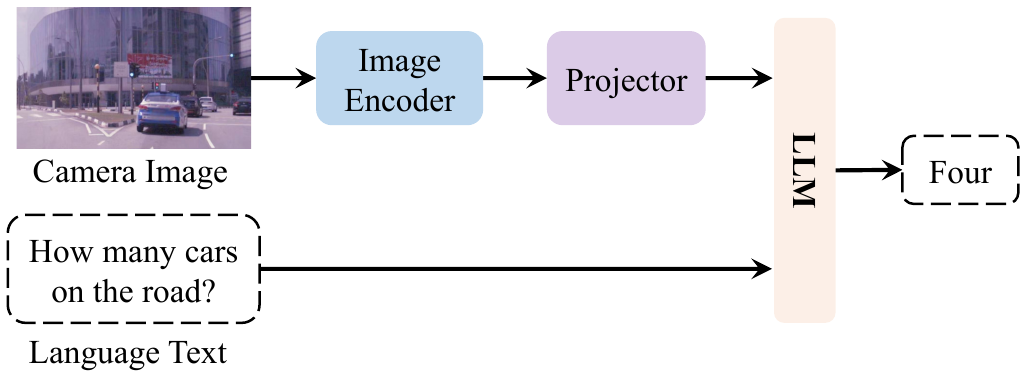}
    \caption{Visual-Language based paradigm.}
    \label{f:Visual-Language-LLM}
\end{figure}

\begin{figure*}[t]
    \centering
    \includegraphics[width=180mm]{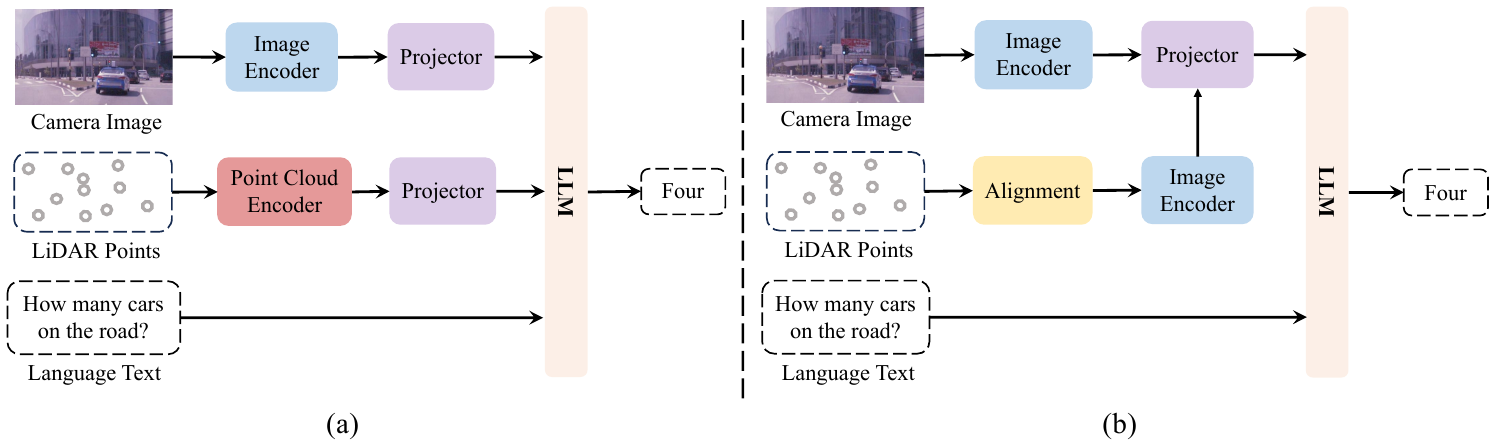}
    \caption{Visual-LiDAR-Language based paradigms. In paradigm~(a), separate encoders are designed for radar and image, and then fused. Paradigm~(b) fuses radar point cloud with image after image modality alignment.}
    \label{f:Visual-LiDAR-LLM1}
\end{figure*}

In recent years, large language model~(LLM) has achieved impressive performance in a variety of tasks. By fusing data from different modalities, multi-modal LLM~(MM-LLM) can perform more complex tasks, such as image captioning, video understanding, and cross-modal retrieval.
Recently, various new datasets have also been designed to advance MM-LLM for embodied AI.
For example, projects like DriveLM~\cite{sima2023drivelm}, OmniDrive~\cite{wang2024omnidrive}, and NuInstruct enhance existing datasets by incorporating large language models (LLMs) to generate question-answer pairs covering perception, reasoning, and planning. 
Additionally, MAPLM~\cite{cao2024maplm} integrates multi-view imagery with LiDAR data to analyze and interpret road surface conditions.
Based on the MM-LLM and those datasets, many studies have been conducted on how to incorporate MM-LLM into MSFP. 
As shown in Table~\ref{t:MM-LLM}, in this section, we mainly review the existing related work from two categories, \textit{i.e.},  \textit{Visual-Language based Methods} and \textit{Visual-LiDAR-Language based Methods}. Pipelines of those two categories are illustrated in Fig.~\ref{f:Visual-Language-LLM} and Fig.~\ref{f:Visual-LiDAR-LLM1}, respectively.

\subsection{Visual-Language based Methods}

Multi-modal large models have demonstrated substantial potential in intelligent perception, with various approaches exploring their capabilities in addressing the complexities of real-world environments. 
X-Driver~\cite{liu2025x} proposes a unified framework leveraging multi-modal large language models with Chain-of-Thought reasoning and autoregressive modeling, achieving superior closed-loop autonomous driving performance, enhanced interpretability. 
Mpdrive~\cite{zhang2025mpdrive} introduces a novel marker-based prompt learning framework, which leverages concise visual markers to represent spatial coordinates and constructs dual-granularity visual prompts, achieving state-of-the-art performance with enhanced spatial perception on tasks requiring advanced spatial understanding.
DriveVLM~\cite{tian2024drivevlm} integrates traditional architectures with MM-LLMs through two distinct branches: one focusing on traditional visual processing and the other leveraging the power of multi-modal transformers for scene understanding. 

Advancements in model design have further enhanced perception and reasoning abilities. Reason2Drive~\cite{nie2025reason2drive} utilizes a prior tokenizer to extract local image features, BEV-InMLLM~\cite{ding2024holistic} incorporates BEV representations for spatial understanding, and OmniDrive~\cite{wang2024omnidrive} integrates 2D pre-trained knowledge with 3D spatial data using Q-Former3D. 
Meanwhile, ELM~\cite{zhou2025embodied} captures temporal information with a time-aware token selection mechanism.  Furthermore, Chen \textit{et al.}~\cite{chen2024driving} propose a novel architecture that fuses an object-level vectorized numeric modalities into any LLMs with a two-stage pre-training and fine-tuning method.

\subsection{Visual-LiDAR-Language based Methods}

Given the limited availability of LiDAR and text data, aligning point cloud features directly with text features presents significant challenges. This difficulty arises because point cloud data, inherently three-dimensional and sparse, lacks the dense, structured nature of textual data. To overcome these challenges, image features are usually leveraged as an intermediary to effectively bridge the gap between text and LiDAR data. 
In this way, the rich visual information available in images can be utilized to facilitate a more seamless integration of these disparate data types.
Along this line,
DriveMLM~\cite{wang2023drivemlm} employs a temporal QFormer to process multi-view images, which can effectively capture temporal dynamics and spatial relationships across different perspectives. This is essential for understanding complex scenes. 

Besides, in multi-modal processing, some methods adopt an indirect approach to handling point cloud data, which converts the point clouds into images to facilitate information extraction. 
This transformation allows for the utilization of established techniques that excel in image processing, thereby enhancing the overall efficiency and effectiveness of MSFP.
For example, MAPLM~\cite{cao2024maplm} projects 3D LiDAR point cloud data into a BEV image, and features are extracted through a visual encoder. 
This approach transforms the 3D data into a 2D representation, making it easier to process with traditional deep learning models designed for image data. By using BEV images, MAPLM bridges the gap between point cloud and image data, enabling the use of powerful visual models like CLIP.
Furthermore, LiDAR-LLM~\cite{yang2025lidar} introduces a novel framework for understanding 3D outdoor scenes by reformulating 3D cognition as a language modeling task, utilizing a Position-Aware Transformer (PAT) and a three-stage training strategy to bridge the 3D-Language modality gap and achieve state-of-the-art performance in tasks like 3D captioning, grounding, and question answering.

\section{Open Challenges and Future Opportunities}
\label{s:open challenges}

In this section, we describe the inherent challenges and the possible future opportunities for MSFP.
Discussions will be carried out in the following from three interconnected levels, \textit{i.e.}, data, model and application.

\subsection{Data Level}

\subsubsection{Data Quality}

A major challenge at the data level is the limited quality and representativeness of existing datasets for multi-sensor fusion. 
Many datasets, such as KITTI, nuScenes, and Waymo Open, suffer from long-tailed distributions~\cite{geiger2012kitti, caesar2020nuscenes, sun2020waymo}. 
This imbalance limits the ability of fusion models to generalize to rare but critical scenarios. 
Moreover, while enhancing sensor diversity, sparser radar point clouds also challenge the feature extraction capabilities of existing models. Compounding these issues are challenges such as missing data, outliers, bias, and drift, as well as the lack of standardized evaluation methods and the limited availability of public datasets~\cite{teh2020sensor}.

Therefore, the development of high-quality datasets is essential to address these challenges. 
Artificial Intelligence Generated Content~(AIGC) techniques could potentially generate synthetic data to fill gaps in real-world datasets, particularly for rare or diverse scenarios, \textit{e.g.}, photorealistic rendering~\cite{yangreal} and diffusion model~\cite{croitoru2023diffusion}. 
To improve the reliability of generated synthetic data, future research could focus on developing automated error detection tools~\cite{ehrlinger2022survey}. Additionally, implementing quantitative quality metrics can help identify issues such as missing data, outliers, and data drift.

\subsubsection{Data Augmentation}         

Data augmentation plays a vital role in improving the robustness and generalizability of MSFP systems.
However, multi-modal data augmentation also introduces unique challenges, particularly in preserving synchronization across different sensor modalities~\cite{wang2021pointaugmenting, rangesh2021predicting}. 
Taking LiDAR-camera fusion as an example, when applying rotations or translations to LiDAR point clouds, equivalent transformations must also be applied to the corresponding camera images to maintain spatial coherence~\cite{xiao2022polarmix}. Any inconsistencies in transformations can disrupt the spatial relationships between modalities, which are crucial for effective sensor fusion. 

To address these challenges, researchers could focus on developing advanced techniques for synchronized data augmentation tailored to MSFP systems. One potential research line is the use of cross-modal geometric constraints to ensure spatial consistency during augmentation~\cite{yuan2023camera}. 
For example, LiDAR point cloud transformations could be coupled with homography transformations in camera images to maintain accurate spatial relationships.
Another promising direction involves AIGC, such as diffusion model~\cite{croitoru2023diffusion}, which could create realistic and synchronized augmentations to model variations such as sensor noise and environmental changes, while ensuring cross-modal consistency.

\subsection{Model Level}

\subsubsection{Effective Fusion Strategies}

At the model level, developing effective fusion strategies involves addressing information loss during the alignment and integration of multi-modal sensor data.  Information loss frequently occurs during the alignment process due to mismatches between sensor modalities, such as cameras and LiDAR, which differ in physical configuration, resolution, and perspective~\cite{meyer2019sensor}. Factors like weather and lighting exacerbate these disparities, making precise synchronization difficult~\cite{zhang2023perception}. Traditional alignment methods, such as projecting point clouds into camera coordinate systems~\cite{zhong2021survey}, often introduce additional errors, leading to suboptimal integration.
Integration processes, such as feature aggregation, further contribute to information loss by compressing sensor data and omitting critical details. For example, transforming point clouds into 2D projections (like BEV or range views) reduces three-dimensional spatial information, such as height, which is vital for capturing scene geometry~\cite{zhao2024bev, chang2023bev}. This compounded loss diminishes model ability to fully utilize the complementary strengths of different sensors, thereby limiting overall performance.

To mitigate these challenges, future research could focus on fusion strategies that jointly optimize alignment precision and data fusion. Multi-representation fusion techniques, such as combining voxel grids, point clouds, and 2D projections, offer a pathway to preserve spatial and semantic richness~\cite{xu2021voxel}. Context-aware approaches leveraging temporal consistency~\cite{zhu2024temporally} and adaptive learning methods~\cite{lubken2024adaptive} can improve alignment by dynamically responding to environmental variations. Additionally, attention mechanisms~\cite{vaswani2017attention} can selectively emphasize critical features of each modality during the integration process. 
Moreover, techniques like self-supervised representation learning~\cite{ericsson2022self} and contrastive learning~\cite{wickstrom2022mixing} offer the potential for capturing and utilizing cross-modal relationships, providing richer and more detailed supervision to refine alignment precision. 
These solutions provide a foundation for reducing information loss and enhancing the robustness of multi-modal fusion systems.

\subsubsection{MM-LLM-based Approach}

MM-LLMs can be utilized to process and fuse data from diverse sources, such as text, images, and sensor outputs, which can greatly enrich the understanding of complex environments~\cite{ye2024mplug, tiantokenize}. 
However, integrating these models in real-world embodied AI applications still poses critical challenges.
A significant challenge involves handling sparse and irregular sensor data, such as LiDAR and radar point clouds.  
The high dimensionality of radar point cloud data necessitates sophisticated preprocessing and feature extraction techniques to transform it into a suitable format for model input. Additionally, radar data is inherently sparse and unstructured, making it more challenging to process compared to more structured data types like images or text~\cite{schumann2021radarscenes, li2020deep}. 
To bridge this gap, future research could investigate hybrid architectures that combine geometric learning techniques, such as graph neural networks or point-based learning models~\cite{svenningsson2021radar,murray2024estimating}, with the multi-modal processing capabilities of MM-LLMs. 

Additionally, external knowledge from MM-LLMs, trained on diverse datasets, may conflict with the specific requirements of embodied AI. 
For instance, an MM-LLM might infer that a busy crosswalk scenario applies universally, suggesting unnecessary stops in environments like highways, where crosswalks are irrelevant. 
Such conflicts between general knowledge and specific contexts could compromise decision-making if not carefully managed or adapted.
To address these challenges, future research could explore leveraging mechanisms like Retrieval-Augmented Generation (RAG) to dynamically adapt external knowledge to the context provided by multi-sensor data~\cite{sharifymoghaddam2024unirag}. Attention mechanisms could further refine this process by emphasizing relevant information and filtering out irrelevant or misleading content. 
These approaches provide a potential pathway for ensuring that external knowledge aligns with the specific, real-time demands of embodied agent systems, enhancing their robustness and reliability.

\subsection{Application Level}

\subsubsection{Real-world Adaptability}

In real-world open environments, conditions are highly variable, including changes in lighting, weather, and traffic patterns. Sudden shifts, such as rain, fog, snow, or the transition from day to night, present significant challenges for multi-sensor fusion systems, which must consistently maintain reliable performance despite these dynamic changes. 
Therefore, effective adaptation to diverse scenarios is essential to ensure the robustness of MSFP systems and prevent failures in challenging conditions.

To enhance real-world adaptability, future research could focus on developing self-adaptive algorithms that can adjust model parameters in response to environmental changes in real-time~\cite{xing2024comprehensive}. 
Techniques such as domain adaptation and online learning~\cite{silva2023adaptive, hoi2021online} could allow models to maintain performance by continuously adapting to new data distributions without retraining from scratch. 
Furthermore, zero-shot learning methods~\cite{wang2019survey, xian2018zero} could be explored to enable models to generalize to unseen scenarios and handle novel environmental conditions without prior specific training, improving their ability to cope with unpredictable real-world situations.

\subsubsection{Explainability}

Explainability is significantly important for MSFP models in embodied AI because it builds trust, ensures transparency, and aids debugging, particularly in safety-critical applications~\cite{de2024building}. 
One of the main challenges lies in identifying the contribution of each sensor modality under varying conditions. 
For instance, LiDAR might play a key role in poor lighting, while cameras may be more effective in clear weather. 
Understanding these contributions, along with how different modalities interact, is difficult because these relationships are often not straightforward, especially in complex real-world scenarios.

Future research could explore context-aware interpretability methods~\cite{wang2023context} to clarify the role of each modality based on environmental conditions and fusion stages. 
For instance, attention-based visualization tools~\cite{jin2022visual, yeh2023attentionviz} might highlight which sensors contributed most under specific scenarios, enhancing transparency in decision-making. 
Additionally, explainable fusion networks could be designed to output modality-specific confidence scores, providing a clear understanding of how each data source impacted the output, especially in critical or ambiguous situations.
Tailoring explanations to modalities and application scenarios can improve user trust and support safer, more effective deployment in real-world environments.

\section{Conclusion}
\label{s:conclusion}

In this survey, we comprehensively reviewed research methods on multi-sensor fusion perception~(MSFP) for embodied AI.
Specifically, first, we introduced the background of MSFP.
Then, we organized and reviewed specific methods from four categories: multi-modal fusion, multi-agent fusion, time-series fusion, and MM-LLM fusion methods. 
Finally, we discussed current challenges and future opportunities.
In retrospect, different from existing surveys that focused on specific fields~(\textit{e.g.}, autonomous driving) or tasks~(\textit{e.g.}, 3D object detection), we organized MSFP research from a task-agnostic perspective, where methods were reported purely from various technical views. 
Therefore, this paper is suitable for a very wide range of researchers to read, including different fields and different tasks.
In the future, we will regularly update this survey online to introduce the latest frontier progress in the field of MSFP.

Due to the limited expertise of the authors, if there are any problems in this paper, please feel free to let us know. 
We will seriously improve the survey in the updated version.



\bibliographystyle{IEEEtran}
\bibliography{ref}

\vfill

\end{document}